\documentclass[twocolumn,amsmath,amssymb,showpacs]{revtex4}
\usepackage{graphicx}

\usepackage{xcolor}
\usepackage{longtable}
\usepackage{times}
\usepackage{dcolumn}
\usepackage{bm}
\usepackage[plainpages = false, pdfpagelabels, 
bookmarks,
bookmarksopen = true,
bookmarksnumbered = true,
linktocpage,
pagebackref=false,
colorlinks = true,
linkcolor = blue,
urlcolor  = blue,
citecolor = blue,
anchorcolor = green,
hyperindex = true,
hyperfigures]
{hyperref} 

\begin{document}

\title {\Large On energy-dependent scaling factor for the charge-changing cross sections of light elements}

\author{Z. \surname{Hasan}$^1$}
\author{M. \surname{Imran}$^{2,3}$}
\author{A. A. \surname{Usmani}$^3$}
\author{Z. A. \surname{Khan}$^3$}\email{zakhan.amu@gmail.com}
\affiliation{$^1$Department of Applied Physics, ZH College of Engineering and Technology, Aligarh Muslim University, Aligarh-202002, India}
\affiliation{$^2$Applied Science \& Humanities Section, University Women's Polytechnic, Aligarh Muslim University, Aligarh-202002, India}
\affiliation{$^3$Department of Physics, Aligarh Muslim University, Aligarh-202002, India}

\begin{abstract}

In this study, the scaling factor for $^{28}$\rm Si + $^{12}$\rm C charge-changing cross sections (CCCSs) at 90-1296 MeV/nucleon has been used as a basis to introduce the energy dependence in the scaling factor for the light elements. To test the extracted scaling factor, we predict the charge-changing cross sections for $^{7,9-12,14}$\rm Be, $^{10-15,17}$\rm B, $^{11-19}$\rm C, $^{13-22}$\rm N, $^{15-24}$\rm O, and $^{18-21,23-26}$\rm F isotopes at 200-991 MeV/nucleon in the framework of the Glauber model. The calculations use descriptions of nuclei in terms of the Slater determinant involving harmonic oscillator single-particle wave functions, which reproduce the charge radii obtained earlier [Phys. Rev. C {\bf 110} (2024) 014623; {\bf 111} (2025) 064601]. It is found that the theoretical results provide quite a satisfactory explanation of the experimental data in all the cases, and the extracted scaling factor shows systematic energy dependence for the isotopes of a given element. In conclusion, we expect that the present way of finding the scaling factor could be successfully used in the analysis of CCCSs for the elements Z $\leq$ 9 at any desired energy in the range 90-1296 MeV/nucleon. This result may further add that the present work provides a practical scheme for predicting the charge-changing cross sections (and inferring proton radii) of \rm Be-\rm  F isotopes, where measurements are scarce.
 
\end{abstract}
\pacs{24.10.Ht, 25.60.Dz, 25.60.-t, 25.70.-z}
\maketitle

\section{Introduction}
\label{sec1}

The production of unstable nuclei using the Radioactive Ion Beam (RIB) facility has added important applications in nuclear physics; studying nuclear structure and properties of exotic nuclei, nuclear astrophysics; investigating nuclear reactions that occur in stars and stellar nucleosynthesis, and nuclear medicine; developing new isotopes for diagnostics and therapies. From the studies of unstable nuclei, it has become clear that the basic properties of nuclear density that are generally applicable to stable nuclei do not hold in unstable nuclei. For the nuclear size, the use of RI beam technology has, however, made it possible to determine the matter radii of a wide range of stable and unstable nuclei by measuring the interaction and reaction cross sections \cite{1}. Such measurements have revealed some unique properties of unstable nuclei such as the existence of thick neutron skins and halos \cite{2,3,4}. These exotic features of unstable nuclei demand to find the proton and neutron distributions, and also the proton and neutron radii as reliable as possible. It is well known that the conventional tools for measuring the charge radii, the electron scattering and isotope-shift (IS) measurements, have been performed particularly for a large section of stable nuclei and for limited unstable (short lived) nuclei around the $\beta$-stability line. However, the availability of unstable nuclei away from the $\beta$-stability line requires involving other measurements to provide the charge radii of such nuclei. As we know, the measurements of the charge-changing cross sections (CCCSs) \cite{5,6,7,8,9,10,11} have been performed with a view that such experiments may be useful to provide information about the charge radii of unstable (exotic) nuclei. It was hoped that the combined study of CCCSs and reaction (interaction) cross sections makes it possible to predict the reliable estimates for the charge and matter radii of exotic nuclei.

Encouraged by the successful use of the Glauber model in explaining the experimental data on interaction cross sections \cite{12}, it was thought that a similar study, involving only the projectile protons, could possibly account for the charge-changing cross sections data. However, the calculations of CCCS using the Glauber model \cite{6,13,14,15,16,17,18} have shown that the projectile neutrons may partially contribute to the CCCS, leaving an unclear picture of the reaction mechanism for CCCS. To overcome the problem, arising due to the presence of projectile neutrons in the study of CCCS, different approaches were adopted \cite{6,13,14,15,16,17,18}, which make it possible to accommodate CCCS within the framework of the Glauber model. As discussed in our recent work \cite{19}, since the charge radii are crucial for obtaining the neutron skin thickness and understanding the halo-like structure of neutron-rich (unstable) nuclei, the approach that involves only the proton distribution and a phenomenological scaling (correction) factor \cite{6,13,14,15,16,19} seems to better suit the very motive of the CCCS measurements, and it is this approach that will be followed in this work also.          
      
To appreciate the purpose of the present work, we divide our proposed plan into two parts: (i) The first part is devoted to make a detailed study of the $^{28}$\rm Si + $^{12}$\rm C charge-changing cross sections in the energy range 90-1296 MeV/nucleon \cite{6,20,21,22,23,24} within the framework of the Glauber model. The Glauber model S matrix has been evaluated using the (a) conventional Glauber model with and without the two-body correlations, (b) optical-limit approximation of the Glauber model, and (c) zero range optical limit approximation of the Glauber model; the reason for calculating the S matrix in different ways will be discussed in Sec. \ref{sec3}. These calculations provide the energy dependence of the scaling factor for $^{28}$\rm Si + $^{12}$\rm C charge-changing cross sections at 90-1296 MeV/nucleon. This scaling factor is also expected to provide satisfactory account for the CCCSs of other isotopes of silicon ($^{25-27}$\rm Si) \cite{25}. (ii) Next, we intended to see how could one make use of the scaling factor for $^{28}$\rm Si + $^{12}$\rm C charge-changing cross sections at 90-1296 MeV/nucleon for studying the CCCSs of other elements, so that one can provide a systematic energy dependence of the scaling factor for the isotopes of the considered elements. The extracted scaling factor is used to predict the charge-changing cross sections for $^{7,9-12,14}$\rm Be, $^{10-15,17}$\rm B, $^{11-19}$\rm C, $^{13-22}$\rm N, $^{15-24}$\rm O, and $^{18-21,23-26}$\rm F isotopes on a $^{12}$\rm C target at 200-991 MeV/nucleon \cite{5,7,8,9,10,11,17,18,26,27} in the framework of the Glauber model. The aforesaid calculations involve descriptions of nuclei in terms of the Slater determinant consisting of the harmonic oscillator single-particle wave functions (hereafter referred to as SDHO densities). The suitability of the extracted scaling factor is judged by comparing our theoretical predictions for CCCSs with the experiment.    
                        
The formulation of the problem is given in  Sec. \ref{sec2}. The numerical results are presented and discussed in Sec. \ref{sec3}. The conclusions are presented in Sec. \ref{sec4}.

\section{Formulation}

\label{sec2}

According to the conventional Glauber model (GM), the elastic S-matrix element, $S_{el}$, for a nucleus-nucleus collision is written as
\begin{equation}
S_{el}(b)= \langle \psi_{T}\psi_{P}\vert\prod^{A}_{i=1}\prod^{B}_{j=1}[1-\Gamma_{NN}(\vec{b}-\vec{s_{i}}+\vec{s^{'}_{j}})]\vert\psi_{P}\psi_{T}\rangle,
\label{eq1}
\end{equation}
where  $\psi_{P}$ ($\psi_{T}$) is the ground-state wave function of projectile (target) nucleus, $A$ $(B)$ is the mass number of target (projectile) nucleus, $\vec{b}$ is the impact parameter vector perpendicular to the incident momentum, $\vec {s_{i}}$ $\vec{(s^{'}_{j}})$ are the projections of target (projectile) nucleon coordinates on the impact parameter plane, and $\Gamma_{NN}({b})$ is the $NN$ profile function, which is related to the $NN$ scattering amplitude $f_{NN}(q)$ as 
follows  
\begin{equation}
\Gamma_{NN}({b})= \frac{1}{2\pi ik}~\int \exp(-i\vec {q}.\vec {b})f_{NN}({q})~d^{2}q, 
\label{eq2}
\end{equation}
where $k$ is the incident nucleon momentum corresponding to the projectile kinetic energy per nucleon, and $\vec{q}$ is the momentum transfer.

Following the approach of Ahmad \cite{28}, the S-matrix element, $S_{el}({b})$, up to two-body correlation (density) term takes the following form:
\begin{equation}
S_{el}({b})\approx S_{0}(b)+S_{2}(b),
\label{eq3}
\end{equation}
where
\begin{equation}
S_{0}({b})= [1-\Gamma^{NN}_{00}(b)]^{AB}, 
\label{eq4}
\end{equation}
and
\begin{gather}
S_{2}({b})=\Bigl\langle\psi_{T}\psi_{P}\Bigl|\frac{1}{2!}[1-\Gamma^{NN}_{00}(b)]^{AB-2}~~~~~~~~~~~~~~~~~~\nonumber\\
~~~~~~~~~~~~~~~~~~\times \sum^{'}_{i_{1},j_{1}}\sum^{'}_{i_{2},j_{2}}\gamma_{i_{1},j_{1}}\gamma_{i_{2},j_{2}}\Bigr|\psi_{P}\psi_{T}\Bigr\rangle,
\label{eq5}
\end{gather}
with 
\begin{equation}
\gamma_{ij}=\Gamma^{NN}_{00}(\vec{b})-\Gamma_{NN}(\vec{b}-\vec{s_{i}}+\vec{s^{'}_{j}}), 
\label{eq6}
\end{equation}
and
\begin{equation}
\Gamma^{NN}_{00}({b})=\int \rho_{T}(\vec r)~\rho_{P}(\vec{r^{'}})~\Gamma_{NN}(\vec b-\vec s +\vec{s^{'}})~d\vec r~d\vec{r^{'}}.
\label{eq7}
\end{equation}
The primes on the summation signs in Eq. (\ref{eq5}) indicate the restriction that two pairs of indices cannot be equal at the same time (for example,
if $i_{1}$ = $i_{2}$ then $j_{1} \neq j_{2}$ and vice versa). The quantities $\rho_{T}$ and $\rho_{P}$ in Eq. (\ref{eq7}) are the (one-body) ground state densities of the target and projectile, respectively. Finally, it is to be pointed out that the distinction between protons and neutrons in both the projectile and target has been included in the expression (3), but only in the uncorrelated part ($S_{0}$). This modification considers the use of different values of the parameters for pp and pn scattering amplitudes and involves different density distributions for protons and neutrons in the colliding nuclei. However, the two-body correlation term ($S_{2}$) involves the average behavior of pp and pn interactions, and considers matter density distributions. More explicitly, the evaluation of $S_{0}({b})$ and $S_{2}({b})$ leads to the following expressions:
\begin{gather}
	S_{0}({b})= [1-\Gamma^{pp}_{00}({b})]^{Z_{P}Z_{T}}[1-\Gamma^{np}_{00}({b})]^{N_{P}Z_{T}}~~~\nonumber\\ 
	~~~~~~~~~~~~~~~~~\times[1-\Gamma^{pn}_{00}({b})]^{Z_{P}N_{T}}[1-\Gamma^{nn}_{00}({b})]^{N_{P}N_{T}},
	\label{eq8}
\end{gather}
\begin{gather}
	S_{2}({b})=-\frac{AB}{8\pi^{2}k^{2}}[1-\Gamma^{NN}_{00}(b)]^{AB-2}[(A-1)(B-1)\nonumber\\
	~~~~~~~~~~~~~~~~~\times(G_{22}(b)-G_{00}^{2}(b))+(B-1)\nonumber\\
	~~~~~~~~~~~~~~~~~\times(G_{21}(b)-G_{00}^{2}(b))+(A-1)\nonumber\\
	~~~~~~\times(G_{12}(b)-G_{00}^{2}(b))],
	\label{eq9}
\end{gather}
with
\begin{equation}
\Gamma^{ij}_{00}({b})=\int \rho^{j}_{T}(\vec r_{j})~\rho^{i}_{P}(\vec{r_{i}^{'}})~\Gamma_{ij}(\vec b-\vec s_{j} +\vec{s_{i}^{'}})~d\vec r_{j}~d\vec{r_{i}^{'}},
\label{eq10}
\end{equation}
where $Z_{T}$ $(Z_{P})$ and $N_{T}$ $(N_{P})$ are the target (projectile) atomic and neutron number, respectively, and each of $i$ and $j$ stands for a proton and a neutron. For the evaluation of $G_{22}(b), G_{21}(b), G_{12}(b)$, and  $G_{00}(b)$, we refer readers to follow the work presented in Ref. \cite{19}.  

\subsection{Charge-changing cross section}

As mentioned earlier, the Glauber model calculations for charge-changing cross section ($\sigma_{cc}$) suggest \cite{6,13,14,15,16,17,18} that in addition to the protons in the projectile, the partial involvement of the projectile neutrons makes it difficult to understand the reaction mechanism for charge-changing cross section. However, keeping in mind the very purpose of CCCS measurements, it was suggested that the calculations of CCCS may be possible by introducing a phenomenological scaling factor \cite{6,13,19,25,29} that can consider the effect of the presence of neutrons in the projectile. Under this approach, the charge-changing cross section is expressed as follows:
\begin{equation}
\sigma_{cc}=\epsilon(E) ~ \sigma^{p}_{cc},
\label{eq11}
\end{equation} 
where $\epsilon(E)$ is the scaling factor which, in general, is energy dependent and is defined as the ratio of the experimental $\sigma_{cc}$ and calculated $\sigma^{p}_{cc}$ values ($\sigma^{exp}_{cc}/\sigma^{p}_{cc}$). The quantity $\sigma^{p}_{cc}$ is the contribution to the charge-changing cross section due to the scattering of only projectile protons. In connection with the scaling factor $\epsilon(E)$, we need to mention that, in our recent works \cite{19,30}, we have adopted a procedure that involves similar value of $\epsilon(E)$ for each of the isotopes of a given element for a particular range of incident energies. However, in this work, we attempt to develop a different prescription, described in the next section, that can provide energy- dependent scaling factor even for a smaller difference in the energies of the isotopes of a given element. Thus, we need to mention only the calculation of $\sigma^{p}_{cc}$ in order to get the results for $\sigma_{cc}$. Following Eq. (\ref{eq3}), the elastic S-matrix element, $S^{p}_{el}$, that considers only the projectile protons, is given by     
\begin{equation}
S^{p}_{el}({b})\approx S^{p}_{0}(b)+S^{p}_{2}(b).
\label{eq12}
\end{equation}
The quantities $S^{p}_{0}(b)$ and $S^{p}_{2}(b)$ can be obtained by putting $N_{P}$ = 0 in the respective expressions for $S_{0}(b)$ (Eq. (\ref{eq8})) and $S_{2}(b)$ (Eq. (\ref{eq9})). This simplification leads to the following expressions for $S^{p}_{0}(b)$ and $S^{p}_{2}(b)$:
\begin{equation}
S^{p}_{0}(b)= [1-\Gamma^{pp}_{00}(b)]^{Z_{P}Z_{T}}[1-\Gamma^{pn}_{00}(b)]^{Z_{P}N_{T}}, 
\label{eq13}
\end{equation}
\begin{gather}
S^{p}_{2}(b)=-\frac{AZ_{P}}{8\pi^{2}k^{2}}(1-\Gamma^{NN}_{00}(b))^{AZ_{P}-2}[(A-1)(Z_{P}-1)\nonumber\\
~~~~~~~~~~~~~~~~~\times (G_{22}(b)-G^{2}_{00}(b))+(Z_{P}-1)\nonumber\\
~~~~~~~~~~~~~~~~~\times (G_{21}(b)-G^{2}_{00}(b))+(A-1)~~\nonumber\\
~~~~\times (G_{12}(b)-G^{2}_{00}(b))],
\label{eq14} 
\end{gather}
The quantities $G_{22}$, $G_{21}$, $G_{12}$, and $G_{00}$ in the above equation assume similar expressions as in Eq. (\ref{eq9}), but now they involve only the projectile proton density instead of projectile matter density. Here, it should be mentioned that the above expressions are valid for conventional Glauber model (GM) calculations. In the case of optical limit approximation of the Glauber model (OLA) \cite{31}, the elastic S-matrix element, $S^{p}_{el}$, takes the form
\begin{equation}
S^{p}_{el}({b})\approx S^{p}_{0}(b),
\label{eq15}
\end{equation}
where
\begin{equation}
S^{p}_{0}(b)= \exp\left[Z_{P}Z_{T}\Gamma^{pp}_{00}(b)+Z_{P}N_{T}\Gamma^{pn}_{00}(b)\right]. 
\label{eq16}
\end{equation}
Further, it should be noted that the calculations for $\sigma_{cc}$ are also performed using the zero range optical limit approximation of the Glauber model (ZROLA). Such calculations also involve similar expression for $S^{p}_{el}({b})$ as in Eq. (\ref{eq15}), but the definition of NN profile function $\Gamma_{NN}(b)$ (Eq. (\ref{eq2})) in ZROLA is not the same as in GM and OLA; this point is further discussed in the next section.
 
With these considerations, the charge-changing cross section for a nucleus-nucleus collision, that considers only the projectile protons, is calculated using the equation
\begin{equation}
\sigma_{cc}^{p}= 2\pi\int \left[1- \vert S^{p} _{el}(b)\vert^{2}\right]b~db.
\label{eq17}
\end{equation}

\section{Results and Discussion}
\label{sec3}
\subsection{Inputs}

(i) Intrinsic proton density distribution: The intrinsic proton density of the projectiles is obtained from the Slater determinant consisting of the harmonic oscillator single-particle wave functions \cite{12}. This density involves the oscillator constant as the only basic input, which, obviously, takes different values for different nuclei, and is related to the corresponding charge radius of a given nucleus. In this work, we require the values of oscillator constant for $^{7,9-12,14}$\rm Be, $^{10-15,17}$\rm B, $^{11-19}$\rm C, $^{13-22}$\rm N, $^{15-24}$\rm O, and $^{18-21,23-26}$\rm F isotopes. For \rm Be, \rm B, \rm C, \rm N, \rm O, and \rm F isotopes, the values of oscillator constant correspond to the charge radii as obtained in Refs. \cite{19,30}, whereas the corresponding values for \rm Si isotopes (given in Table \ref{tab1}) are obtained using the charge radii as calculated using the deformed relativistic Hartree-Bogoliubov theory in continuum (DRHBc) \cite{32}. For the target $^{12}$\rm C nucleus, we involve the charge density as obtained from the electron-scattering experiment \cite{33} and assume the neutron and proton densities to be the same. 
  
  \begin{table}	
 	\caption{The oscillator constant, $\rm \alpha_p^2$, reproduces the projectile charge radius ($\rm r_{p}$) as calculated using the deformed relativistic Hartree-Bogoliubov theory in continuum (DRHBc) \cite{32}.}  
 	\label{tab1}
 	\begin{tabular}{ccc}
 		\\
 		\hline
 		\hline
 		&&\\
 		$\rm Projectile$&$\rm \alpha_p^2$&$\rm r_{p}$    \\
 		 	         	&$\rm (fm^{-2})$ &\rm (fm)                                    \\
 		\\                                              	
 		\hline
 		\hline
 		&&\\
 		$\rm ^{25}Si$     &0.2689 &3.184               \\
 		$\rm ^{26}Si$     &0.2887 &3.074                            \\
 		$\rm ^{27}Si$     &0.2837  &3.102                             \\
 		$\rm ^{28}Si$     &0.2845  &3.098                           \\
 		
 		\hline
 		\hline
 	\end{tabular}
 \end{table}
 

(ii) Nucleon-nucleon (NN) amplitude: The NN scattering amplitude $f_{NN}(q)$ is usually parametrized in the form \cite{34}
\begin{equation}
f_{NN}(\vec q)=\frac{ik\sigma_{NN}}{4\pi}
(1-i\rho_{NN})\exp\left(-\beta_{NN}q^{2}/2\right).
\label{eq18}
\end{equation}
It consists of three parameters; $\sigma_{NN}$, $\rho_{NN}$, and $\beta_{NN}$. The values of these parameters at the desired energies (90-1296 MeV) have been obtained by a linear interpolation and extrapolation of their values given in Ref. \cite{34}. From calculation point of view, it should be mentioned that the GM and OLA involve the above form of the NN amplitude, which gives the following expression for the NN profile function $\Gamma_{NN}({b})$  
\begin{equation}
\Gamma_{NN}({b})= \frac{\sigma_{NN}}{4\pi \beta_{NN}}~(1-i\rho_{NN})\exp\left[-b^{2}/(2\beta_{NN})\right].
\label{eq19}
\end{equation}
Whereas, in the ZROLA, the NN amplitude (Eq. (\ref{eq18})) assumes $\beta_{NN}$= 0, and one gets
\begin{equation}
\Gamma_{NN}({b})= \frac{\sigma_{NN}(1-i\rho_{NN})}{2}~\delta(b).
\label{eq20}
\end{equation}
\subsection{Calculation}

(i) $^{28}$\rm Si + $^{12}$\rm C charge-changing cross sections at 90-1296 MeV/nucleon: To start with, it is important to mention that, generally, the Glauber model calculations for a nucleus-nucleus charge-changing cross section (CCCS), using the scaling factor approach, involved the ZROLA to calculate the NN profile function $\Gamma_{NN}({b})$, and assessed the importance of the presence of projectile neutrons in explaining the CCCS data. In this connection, however, we feel that the use of ZROLA alone for understanding the magnitude of the contribution of projectile neutrons to CCCS may be misleading, as there are other ways that can be used to find the scaling factor in the framework of the Glauber model, and it is not necessary that all the methods would provide similar behavior of the energy dependence of the scaling factor. It is in this context that the Glauber model S matrix $S^{p}_{el}({b})$, needed to obtain the scaling factor, has been evaluated using the (a) GM with and without the two-body correlations, (b) OLA, and (c) ZROLA. Our aim is to see how far the value of the scaling factor differs when obtained using the different ways for calculating $S^{p}_{el}$, and what effects we observe on the results for CCCS.
 \begin{figure}
	\begin{center}
 	\includegraphics[height=8.9cm, width=8.9cm]{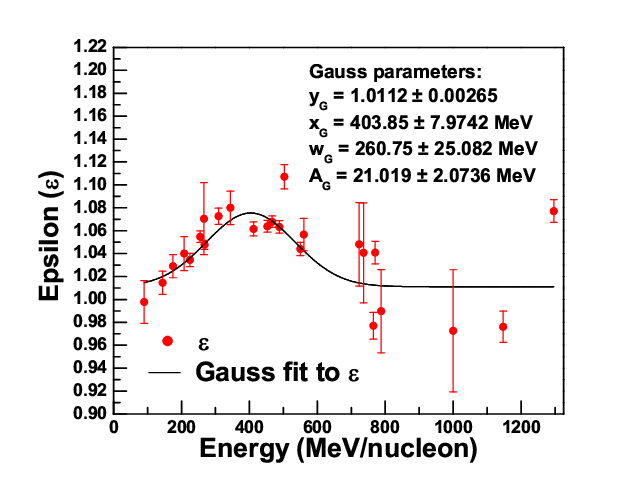}
		\caption{Energy dependence of the scaling factor $\epsilon(E)$ (=$\sigma_{cc}^{exp}/\sigma_{cc}^{p}$) for the charge-changing cross sections (CCCSs) of $^{28}$\rm Si on $^{12}$\rm C at 90-1296 MeV/nucleon. $\sigma_{cc}^{exp}$ is the experimental CCCS and $\sigma_{cc}^{p}$ provides the contribution to CCCS due to projectile protons. $\sigma_{cc}^{p}$ is calculated using the GM with two-body correlations.}  
		\label{fig1}
	\end{center}
\end{figure}

 \begin{figure}
	\begin{center}
				\includegraphics[height=8.9cm, width=8.9cm]{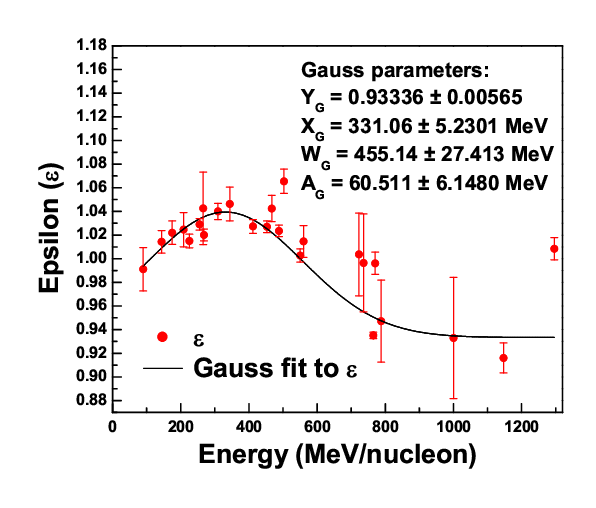}
						\caption{Same as in Fig. \ref{fig1}, but the $\sigma_{cc}^{p}$ is calculated using the GM without two-body correlations.}  
		\label{fig2}
	\end{center}
\end{figure}
 \begin{figure}
	\begin{center}
						\includegraphics[height=8.9cm, width=8.9cm]{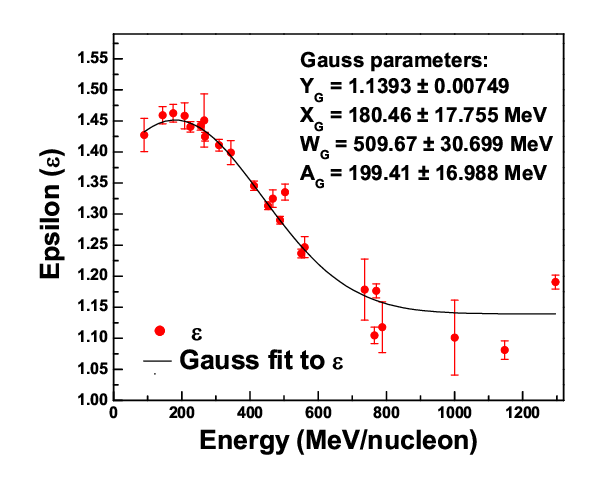}
						\caption{Same as in  Fig. \ref{fig1}, but the $\sigma_{cc}^{p}$ is calculated using the optical limit approximation of the Glauber model (OLA).}  
		\label{fig3}
	\end{center}
\end{figure}
 \begin{figure}
	\begin{center}
		
	\includegraphics[height=8.9cm, width=8.9cm]{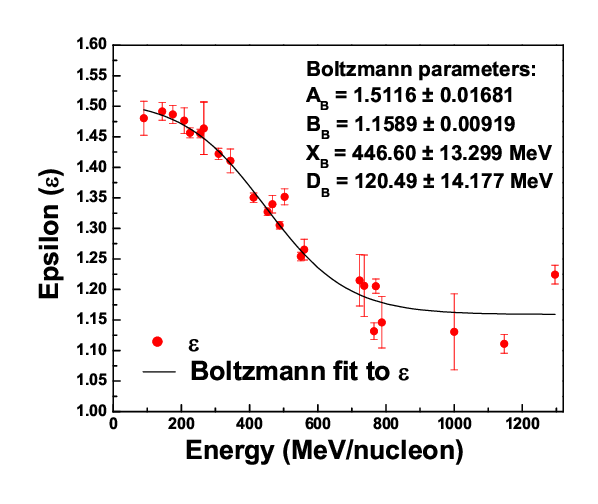}
				\caption{Same as in  Fig. \ref{fig1}, but the $\sigma_{cc}^{p}$ is calculated using the zero range optical limit approximation of the Glauber model (ZROLA).}  
		\label{fig4}
	\end{center}
\end{figure}
 \begin{table*}
 	\caption{$\rm \sigma_{cc}$ gives the calculated charge-changing cross section for $^{25-28}$\rm Si on $^{12}$\rm C at energy E using the GM with two-body correlations, GM without two-body correlations, OLA, and ZROLA. The experimental charge-changing cross section is shown by $\rm \sigma_{cc}^{exp}$. $(\Delta\rm \sigma_{cc})\%$ gives the maximum  percentage difference between $\rm \sigma_{cc}^{exp}$ and $\rm \sigma_{cc}$.}  
 	\label{tab2}
 	\begin{ruledtabular}
 		\begin{tabular}{cccccccc}
 			\\
 			Projectile& E/A (MeV) & $\rm \sigma_{cc}^{exp} (mb)$& \multicolumn{4}{c}{$\rm \sigma_{cc} (mb)$}&$(\Delta\rm \sigma_{cc})\%$ \\
 			\cline{4-7}\\
 			 			&&  \cite{25}&GM with two-body &GM without two-body &OLA & ZROLA & 
 			\\
 			&&&correlations &correlations &&	\\
 				&                    & &&&                                               \\	
 			\hline
 			\\
 			$\rm ^{25}Si$   &268&1130.9$\pm$4.8&1125.9&1141.3&1133.6&1134.1&$\sim$ 0.9    \\
 			$\rm ^{26}Si$   &263&1118.4$\pm$4.5&1092.8&1107.8&1104.6&1104.5&$\sim$ 2.2     \\
 			$\rm ^{27}Si$  &261&1128.9$\pm$4.2&1101.2 &1116.9&1111.7&1111.7&$\sim$ 2.4                        \\
 			$\rm ^{28}Si$  &268&1106.0$\pm$5.0&1099.6&1115.3&1112.2&1112.2&$\sim$ 0.8                          \\
 		\end{tabular}
 	\end{ruledtabular}
 \end{table*}
 
 \begin{figure}
	\begin{center}
		 		\includegraphics[height=6.cm, width=8.9cm]{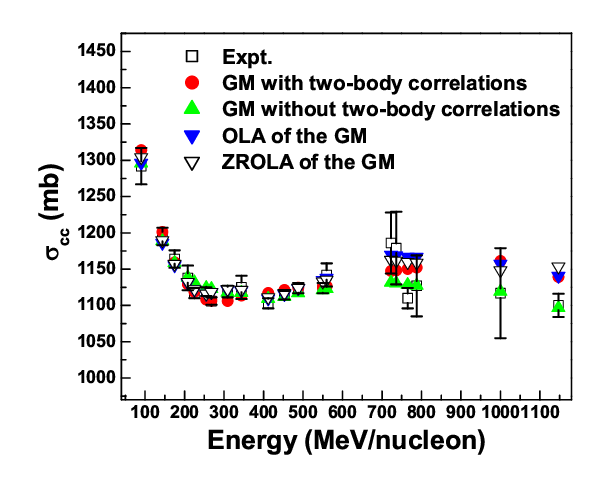}
 				\caption{Charge-changing cross section ($\sigma_{cc}$) for $^{28}$\rm Si on $^{12}$\rm C using the average scaling factor $\epsilon_{avg}(E)$ obtained from Eqs. (\ref{eq21}) and (\ref{eq22}). The filled circles, filled triangles, filled inverted triangles, and open inverted triangles correspond, respectively, to the predicted $\sigma_{cc}$ using the GM with two-body correlations, GM without two-body correlations, OLA, and ZROLA. The experimental data are taken from Refs. \cite{6,20,21,22,23,24}.}  
		\label{fig5}
	\end{center}
\end{figure}
Figs. \ref{fig1}-\ref{fig4} show, respectively, the behavior of the scaling factor obtained using the GM with two-body correlations, GM without two-body correlations, OLA, and ZROLA for the $^{28}$\rm Si + $^{12}$\rm C charge-changing cross sections at 90-1296 MeV/nucleon; the analysis of these scaling factors is also shown by a continuous line in each figure, which provides the average scaling factor, $\epsilon_{avg}(E)$, in each case. The analysis of the data in Figs. \ref{fig1}-\ref{fig3} involves the Gauss fit, whereas the Boltzmann fit is performed in Fig. \ref{fig4}; 
 \begin{figure*}
	\begin{center}
				\includegraphics[height=14.0cm, width=14.0cm]{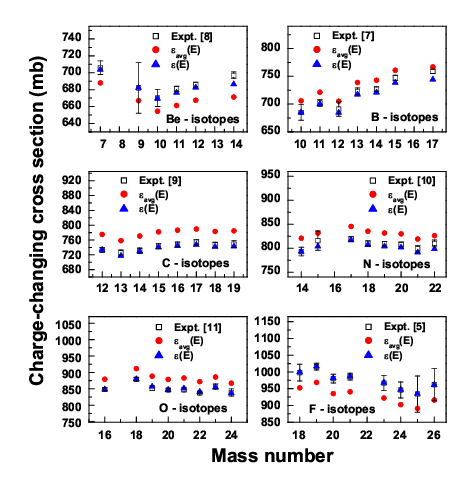}
				\caption{Charge-changing cross sections (CCCSs) for  $\rm Be$, $\rm B$, $\rm C$, $\rm N$, $\rm O$, and $\rm F$ isotopes  on $^{12}$\rm C at energies given in Table \ref{tab3}. The filled circles and triangles represent, respectively, the calculated values of CCCSs involving (i) the average scaling factor $\epsilon_{avg}(E)$ obtained from the analysis of $^{28}$\rm Si + $^{12}$\rm C charge-changing cross sections at 90-1296 MeV/nucleon (Eq. (\ref{eq21}); Fig. \ref{fig1}), and (ii) the scaling factor $\epsilon(E)$ obtained for the said elements (using Eq. (\ref{eq23})); the values of  $\epsilon_{avg}(E)$ and  $\epsilon(E)$ are reported in Table \ref{tab3}.} 
		\label{fig6}
	\end{center}
\end{figure*}

Gauss fit:
\begin{equation}
\epsilon_{avg}(E)= Y_{G} +  \frac{A_{G}}{W_{G}\sqrt{\pi/2}}~\exp\left[-2\{(E-X_{G})/W_{G}\}^{2}\right],
\label{eq21}
\end{equation}

Boltzmann fit:
\begin{equation}
\epsilon_{avg}(E)= \frac{(A_{B}-B_{B})}{1 + \exp\left[(E-X_{B})/D_{B}\right]} + B_{B},
\label{eq22}
\end{equation}
where the energy E is in MeV, and the values of the parameters; $Y_{G}$, $A_{G}$, $W_{G}$, $X_{G}$, $A_{B}$, $B_{B}$, $X_{B}$, $D_{B}$ along with their uncertainties are given in the respective Figs. \ref{fig1}-\ref{fig4}. As expected, Figs. \ref{fig1}-\ref{fig4} show that the GM with and without two-body correlations, OLA, and ZROLA provide different estimates for the scaling factor, showing that one cannot judge about the absolute magnitude of the effect of the projectile neutrons to CCCS, rather one may only suggests the need to include the effect of the projectile neutrons in the calculation of CCCS. Here, it is to be noted that, in the present work, our calculations consider the average value of $\epsilon$ ($\epsilon_{avg}(E)$) without involving uncertainty in the parameters; $Y_{G}$, $A_{G}$, $W_{G}$, $X_{G}$, $A_{B}$, $B_{B}$, $X_{B}$, $D_{B}$. Further, these calculations also do not include uncertainty in the proton radii \cite{19,30}, and, hence, the predicted values of CCCSs will be compared with the central values of the corresponding experimental data. However, following Ref. \cite{19}, it may be added that consideration of the uncertainties in parameters ($Y_{G}$, $A_{G}$, $W_{G}$, $X_{G}$, $A_{B}$, $B_{B}$, $X_{B}$, $D_{B}$) as well as in proton radii would simply introduce uncertainties in the final calculated CCCSs, without altering the conclusions of the present work.  
 
 Next, we consider it worthwhile to see how well the average scaling factor $\epsilon_{avg}(E)$ (Eqs. (\ref{eq21}) and (\ref{eq22}); Figs. \ref{fig1}-\ref{fig4}) accounts for the $^{28}$\rm Si + $^{12}$\rm C charge-changing cross sections at 90-1296 MeV/nucleon, and what could be said about the different methods; GM, OLA, and ZROLA used in the study of CCCS. The results of such calculations are presented in Fig.  \ref{fig5}. As expected, it is found that the GM with and without two-body correlations, OLA, and ZROLA provide equivalent results using the average scaling factor $\epsilon_{avg}(E)$, and all the predicted results lie within experimental errors. Further, in order to test the average scaling factor, we have calculated the CCCSs for other isotopes of silicon ($^{25-27}$\rm Si) (including the $^{28}$\rm Si) on $^{12}$\rm C at $\sim$260 MeV/nucleon \cite{25}. The results are found to provide fairly good agreement between theory and experiment in all the cases (see Table \ref{tab2}); the maximum difference between the calculated and experimental values is about 2$\%$, showing that the average scaling factor $\epsilon_{avg}(E)$ works well in the analysis of CCCS for the available isotopes of \rm Si. Thus, despite the fact that different approximations (GM, OLA, ZROLA) lead to noticeably different behaviors of $\epsilon(E)$, the findings in Table \ref{tab2} and Fig. \ref{fig5} show that the final $\rm \sigma_{cc}$ values are quite similar. This result may indicate that the scaling factor compensates for different estimates for $\rm \sigma_{cc}^{p}$ (Eq. (\ref{eq11})) obtained using different model approximations. In conclusion, we arrive at the result that one can use any one of the Glauber model approaches; GM, OLA, and ZROLA for obtaining the scaling factor, used in the analysis of CCCS.
 
 \begin{table}
 		\caption{$\epsilon(E)$ (= $\epsilon_{avg}(E)$+$\Delta\epsilon$) is the extracted scaling factor (Eq. (\ref{eq23})) used in the analysis of the charge-changing cross section (CCCS) for the considered projectile at energy E. The average scaling factor $\epsilon_{avg}(E)$ is obtained from the analysis of CCCSs for  $^{28}$\rm Si + $^{12}$\rm C at 90-1296 MeV/nucleon (Eq. (\ref{eq21}); Fig. \ref{fig1}); $\Delta\epsilon$ is the correction to $\epsilon_{avg}(E)$.} 
 	\label{tab3}
 	\begin{tabular}{ccccccccc}		
 		\\
 		\hline
 		\hline
 		\\
 		Projectile ~~~&$\rm E/A \rm (MeV)$     ~~~ &$\epsilon_{avg}(E)$~~~~~&$\Delta\epsilon$ ~~~ &$\epsilon(E)$ \\
 	               &       &                   &                  &                           \\
 		\hline     
 		\\
 		$\rm ^{7}Be$    &$772$           &1.0124&0.0230           &1.0354                             \\     
 		$\rm ^{9}Be$    &$921$           &1.0112&           &1.0342                             \\
 		$\rm ^{10}Be$   &$946$           &1.0112&           &1.0342                             \\
 		$\rm ^{11}Be$   &$962$           &1.0112&           &1.0342                             \\
 		$\rm ^{12}Be$   &$925$           &1.0112&           &1.0342                             \\
 		$\rm ^{14}Be$   &$833$           &1.0114&           &1.0344                             \\
 		\\
 		$\rm ^{10}B$    &$925$           &1.0112&-0.0297           &0.9815                             \\     
 		$\rm ^{11}B$    &$932$           &1.0112&           &0.9815                             \\
 		$\rm ^{12}B$    &$991$           &1.0112&           &0.9815                             \\
 		$\rm ^{13}B$    &$897$           &1.0112&           &0.9815                             \\
 		$\rm ^{14}B$    &$926$           &1.0112&           &0.9815                             \\
 		$\rm ^{15}B$    &$920$           &1.0112&           &0.9815                             \\
 		$\rm ^{17}B$    &$862$           &1.0113&           &0.9816                             \\
 		\\
 		$\rm ^{12}C$    &$937$           &1.0112&-0.0541           &0.9571                             \\     
 		$\rm ^{13}C$    &$828$           &1.0115&           &0.9574                             \\
 		$\rm ^{14}C$    &$900$           &1.0112&           &0.9571                             \\
 		$\rm ^{15}C$    &$907$           &1.0112&           &0.9571                             \\
 		$\rm ^{16}C$    &$907$           &1.0112&           &0.9571                             \\
 		$\rm ^{17}C$    &$979$           &1.0227&           &0.9571                             \\
 		$\rm ^{18}C$    &$895$           &1.0112&           &0.9571                             \\
 		$\rm ^{19}C$    &$895$           &1.0112&           &0.9571                             \\
 		\\
 		$\rm ^{14}N$    &$932$           &1.0112&-0.0342           &0.9770                             \\     
 		$\rm ^{15}N$    &$776$           &1.0123&           &0.9781                             \\
 		$\rm ^{17}N$    &$938$           &1.0112&           &0.9770                             \\
 		$\rm ^{18}N$    &$927$           &1.0112&           &0.9770                             \\
 		$\rm ^{19}N$    &$896$           &1.0112&           &0.9770                             \\
 		$\rm ^{20}N$    &$891$           &1.0112&           &0.9770                             \\
 		$\rm ^{21}N$    &$876$           &1.0113&           &0.9771                             \\
 		$\rm ^{22}N$    &$851$           &1.0113&           &0.9771                             \\
 		\\
 		$\rm ^{16}O$    &$857$           &1.0113&-0.0361           &0.9752                             \\     
 		$\rm ^{18}O$    &$872$           &1.0113&           &0.9752                             \\
 		$\rm ^{19}O$    &$956$           &1.0112&           &0.9751                             \\
 		$\rm ^{20}O$    &$880$           &1.0112&           &0.9751                             \\
 		$\rm ^{21}O$    &$937$           &1.0112&           &0.9751                             \\
 		$\rm ^{22}O$    &$937$           &1.0112&           &0.9751                             \\
 		$\rm ^{23}O$    &$871$           &1.0113&           &0.9752                             \\
 		$\rm ^{24}O$    &$866$           &1.0113&           &0.9752                             \\
 		\\
 		$\rm ^{19}F$    &$930$           &1.0112&0.0494           &1.0606                             \\			
 		\hline
 		\hline
 		\\	
 	\end{tabular}
 \end{table}
 \begin{table*}
 	 	\caption{Charge-changing cross sections (CCCSs) for C, N, and O isotopes on $^{12}$\rm C at energy $\sim$200-300 MeV/nucleon. $\sigma_{cc}^{avg}$ gives the value of CCCS, obtained using the average scaling factor $\epsilon_{avg}$(E) for $^{28}$\rm Si + $^{12}$\rm C CCCS at energy E (Eq. (\ref{eq21}); Fig. \ref{fig1}). $\sigma_{cc}$ is the calculated value of CCCS at energy E, obtained using the scaling factor $\epsilon$(E) (= $\epsilon_{avg}(E)$+$\Delta\epsilon$) as extracted from Eq. (\ref{eq23}); $\Delta\epsilon$ is the correction to $\epsilon_{avg}(E)$. The experimental CCCS is shown by $\rm \sigma_{cc}^{exp}$. }
 	\label{tab4}
 	\begin{tabular}{cccccccc}	
 		\\
 		\hline
 		\hline
 		&&&&&&\\
 		$\rm Projectile$  &$\rm E/A$&$\rm \epsilon_{avg}$(E)&$\rm \sigma_{cc}^{avg}$&$\Delta\epsilon$&$\rm \epsilon $(E) &$\rm \sigma_{cc}$  &$\rm  \sigma_{cc}^{exp}$    \\
 		&$\rm (MeV)$&                     &$\rm (mb)$             &                &   &$\rm (mb)$               &$\rm (mb)$                                            \\
 		&         &                       &                       &                &   &                         & \\		                         
 		\hline
 		&&&&&&&\\
 		 		$E \sim200 MeV/nucleon$ $\cite{27}$ &&&&&&&                            \\  
 		&&&&&&&\\                        
 		$\rm ^{12}C$   &228      &1.0371    &736.9&-0.0192    &1.0179            &723.2           &723$\pm$23           \\     
 		$\rm ^{13}C$   &231      &1.0371    &722.5&    &1.0187            &709.1           &720$\pm$25            \\
 		$\rm ^{14}C$   &234      &1.0387    &731.5&    &1.0195            &718.0           &707$\pm$13             \\
 		$\rm ^{15}C$   &236      &1.0392    &741.9&    &1.0200            &728.3           &749$\pm$19              \\
 		$\rm ^{16}C$   &237      &1.0395    &745.5&    &1.0203            &731.7           &738$\pm$17               \\
 		&&&&&&&\\
 		$\rm ^{14}N$   &223      &1.0357    &785.4&0.0112    &1.0469            &793.9           &843$\pm$32              \\     
 		$\rm ^{15}N$   &226      &1.0365    &799.7&    &1.0477            &808.4           &808$\pm$15               \\
 		$\rm ^{17}N$   &236      &1.0392    &803.5&    &1.0504            &812.2           &809$\pm$17                \\
 		&&&&&&&\\
 		$\rm ^{16}O$   &219      &1.0347    &847.8&0.0173    &1.0520            &861.9           &862$\pm$17                 \\
 		&&&&&&&\\
 		$E \sim300 MeV/nucleon$ $\cite{26}$ &&&&&&&\\
 		&&&&&&&\\
 		$\rm ^{11}C$   &319      &1.0632    &706.5&0.0120     &1.0752           &714.5            &716$\pm$20                  \\
 		$\rm ^{12}C$   &294      &1.0563    &723.1&     &1.0683           &731.3            &731$\pm$52                  \\
 		$\rm ^{13}C$   &322      &1.0640    &713.0&     &1.0760           &721.1            &729$\pm$22                  \\	
 		$\rm ^{14}C$   &339      &1.0680    &725.1&     &1.0800           &733.3            &732$\pm$22                   \\
 		$\rm ^{15}C$   &327      &1.0652    &734.5&     &1.0772           &742.8            &758$\pm$56                   \\
 		&&&&&&&\\
 		$\rm ^{13}N$   &310      &1.0608    &720.9&0.0385     &1.0993           &747.0            &752$\pm$35                  \\	
 		$\rm ^{14}N$   &289      &1.0548    &770.2&     &1.0933           &798.4            &878$\pm$77                  \\
 		$\rm ^{15}N$   &315      &1.0622    &786.9&     &1.1007           &815.4            &815$\pm$11                 \\
 		$\rm ^{16}N$   &322      &1.0640    &782.9&     &1.1025           &811.3            &813$\pm$9                 \\
 		$\rm ^{17}N$   &328      &1.0655    &796.6&     &1.1040           &825.4            &790$\pm$11                \\
 		&&&&&&&\\
 		$\rm ^{15}O$   &301      &1.0583    &842.1&0.0463     &1.1046           &878.9            &880$\pm$18                \\
 		$\rm ^{17}O$   &308      &1.0602    &829.8&     &1.1065           &866.0            &866$\pm$11                \\
 		$\rm ^{18}O$   &368      &1.0731    &868.4&     &1.1194           &905.6            &887$\pm$39                 \\
 		
 		\hline
 		\hline
 		\\
 	\end{tabular}
 \end{table*}
 
(ii) Toward finding scaling factor for the charge-changing cross sections of Be, B, C, N, O, and F elements on a $^{12}$\rm C target at 772-991 MeV/nucleon \cite{5,7,8,9,10,11}: In the analysis of $^{28}$\rm Si + $^{12}$\rm C charge-changing cross sections at 90-1296 MeV/nucleon, we found that although the energy dependence of the scaling factor obtained using GM with and without two-body correlations, OLA, and ZROLA is altogether different (Figs. \ref{fig1}-\ref{fig4}), but the corresponding CCCSs (Fig. \ref{fig5}) are not significantly sensitive to such variation in the scaling factor. This result further suggests that any change in the calculation of $\sigma^{p}_{cc}$, arising due to use of the GM (with and without two-body correlations), OLA, and ZROLA, is being accommodated by the corresponding value of the scaling factor $\epsilon_{avg}(E)$, without introducing a significant change in the predicted behavior of $\sigma_{cc}$. Keeping this in mind, we may, therefore, adopt any one of the methods; GM, OLA, and ZROLA in the calculation of $\sigma^{p}_{cc}$; we have involved the GM up to two-body correlations in the following discussion. However, before proceeding, let us shed some light on the behavior of the average scaling factor $\epsilon_{avg}(E)$ for $^{28}$\rm Si + $^{12}$\rm C charge-changing cross sections at 90-1296 MeV/nucleon, observed in Figs. \ref{fig1}-\ref{fig4}: As noticed, the average scaling factor $\epsilon_{avg}(E)$ is almost energy independent, in all the situations, beyond 700 MeV/nucleon, whereas its energy dependence is clearly visible at relatively lower energies. We expect to have similar energy dependence for the scaling factor for other elements too. With this expectation, we shall  consider the isotopes of a given element in two groups: In the first group, we shall have those isotopes whose energy/nucleon is more than 700 MeV, whereas the second group involves the isotopes having energies close to each other, but less than 700 MeV/nucleon.         

First of all, we would like to assess how far the results of $^{28}$\rm Si + $^{12}$\rm C charge-changing cross sections at 90-1296 MeV/nucleon account for the CCCSs for other elements. For this, we calculate the charge-changing cross sections for $^{7,9-12,14}$\rm Be, $^{10-15,17}$\rm B, $^{12-19}$\rm C, $^{14,15,,17-22}$\rm N, $^{16,18-24}$\rm O, and $^{18-21,23-26}$\rm F isotopes on a $^{12}$\rm C target at 772-991 MeV/nucleon, involving the average scaling factor $\epsilon_{avg}(E)$ (Eq. (\ref{eq21}); Fig. \ref{fig1}). The results of these calculations are given in Fig. \ref{fig6}. It is found that the predicted CCCSs show large deviations from the experimental data in all the cases, suggesting the need to modify the average scaling factor, obtained in the analysis of $^{28}$\rm Si + $^{12}$\rm C charge-changing cross sections (Fig. \ref{fig1}). So, our next aim is to see how could one make use of the average scaling factor (Fig. \ref{fig1}), so that one can accommodate the CCCS data of other elements and provide the energy dependence of the scaling factor for the isotopes of a given element. To achieve this, we propose to adopt the following procedure. To start with, we choose a stable isotope, referred to as a 'test' isotope, among the isotopes of a given element, and write the required scaling factor for this isotope as
\begin{equation} 
\epsilon(E) = \epsilon_{avg}(E) + \Delta\epsilon, 
\label{eq23}                          
\end{equation}
where $\epsilon_{avg}(E)$ is the average scaling factor for $^{28}$\rm Si + $^{12}$\rm C charge-changing cross sections at energy E (Eq. (\ref{eq21}); Fig. \ref{fig1}) and $\Delta\epsilon$ is the correction to $\epsilon_{avg}(E)$ which is treated as an adjustable parameter to reproduce the CCCS for the 'test' isotope. The value of $\Delta\epsilon$, so obtained, is then used to predict the CCCS for all other isotopes of the considered element at respective energies. Here, it may be mentioned that $^{9}$\rm Be, $^{10}$\rm B, $^{12}$\rm C, $^{14}$\rm N, $^{16}$\rm O, and $^{19}$\rm F isotopes are considered as the 'test' isotopes for the respective elements. The extracted values of $\epsilon(E)$ for the isotopes of Be, B, C, N, O, and F are given in Table \ref{tab3} and the corresponding predicted CCCSs are presented in Fig. \ref{fig6}. With the values of the scaling factor $\epsilon(E)$ (given in Table \ref{tab3}), it is found that the predicted CCCSs now provide quite a satisfactory explanation of the experimental data in all the cases. Regarding the scaling factor $\epsilon(E)$, we find (Table \ref{tab3}) that it has nearly similar values for all the isotopes of a given element, beyond 700 MeV/nucleon; this result follows the behavior similar to that observed for the average scaling factor $\epsilon_{avg}(E)$ in the study of $^{28}$\rm Si + $^{12}$\rm C charge-changing cross sections (Fig. \ref{fig1}). For \rm F isotopes, we have provided the value of the scaling factor $\epsilon(E)$ for $^{19}$\rm F only, as the incident energy for the remaining isotopes of \rm F is same and hence each \rm F isotope uses similar value of the scaling factor in the analysis of CCCS.

(iii) Toward finding scaling factor for the charge-changing cross sections of C, N, and O elements on a $^{12}$\rm C target at $\sim$200 and 300 MeV/nucleon \cite{26,27}: In this part, we consider the analysis of the experimental data for CCCSs of (a) $^{12-16}$\rm C, $^{14,15,17}$\rm N, and $^{16}$\rm O isotopes at $\sim$200 MeV/nucleon \cite{27}, and (b) $^{11-15}$\rm C, $^{13-17}$\rm N, and $^{15,17,18}$\rm O isotopes at $\sim$300 MeV/nucleon \cite{26}. Following the procedure, outlined above, we consider $^{12}$\rm C and $^{15}$\rm N at $\sim$200 and 300 MeV/nucleon, $^{16}$\rm O at $\sim$200 MeV/nucleon, and $^{17}$\rm O at $\sim$300 MeV/nucleon as the 'test' nuclei. Here, it may be noted that we have chosen $^{17}$\rm O as a 'test' nucleus instead of $^{16}$\rm O at $\sim$300 MeV/nucleon, because no experimental CCCS is available for $^{16}$\rm O at this energy; in the case of nitrogen, the relatively smaller error in the experimental CCCS for $^{15}$\rm N and the overall trend of the experimental CCCSs for \rm N isotopes favors to choose $^{15}$\rm N as a 'test' isotope instead of the $^{14}$\rm N one.  Next, taking $\epsilon_{avg}(E)$ from Fig. \ref{fig1}, Eq. (\ref{eq23}) is employed to find the values of $\Delta\epsilon$ that reproduce the CCCSs for the 'test' nuclei on a $^{12}$\rm C target at the respective energies. These values of $\Delta\epsilon$ are then used to predict the CCCSs for the remaining isotopes. The extracted values of the scaling factor $\epsilon(E)$ and the corresponding (calculated) CCCSs for the considered isotopes at $\sim$200 and 300 MeV/nucleon are given in Table \ref{tab4}. For completeness, Table \ref{tab4} also contains the values of CCCS, obtained using the average scaling factor $\epsilon_{avg}(E)$ (Fig. \ref{fig1}). Here, we notice that, like the $^{28}$\rm Si +  $^{12}$\rm C case (Fig. \ref{fig1}), the energy dependence of the scaling factor  $\epsilon(E)$ is also reflected in the analysis of the CCCSs for C, N, and O isotopes at $\sim$200 and 300 MeV/nucleon. Regarding the calculated values of the CCCSs ($\sigma_{cc}$) (Table \ref{tab4}), it is found that we get satisfactory account of the data for the isotopes of C, N, and O elements on a $^{12}$\rm C target at $\sim$200 and 300 MeV/nucleon. Thus, we conclude that the study of the scaling factor for $^{28}$\rm Si + $^{12}$\rm C charge-changing cross sections at 90-1296 MeV/nucleon could be taken as a starting point to  find the systematic energy-dependent scaling factor for relatively lighter elements and one is able to provide satisfactory account of the CCCS data for Be, B, C, N, O, and F isotopes on a $^{12}$\rm C target at 200-991 MeV/nucleon. 

Finally, we consider it important to comment on the choice of a 'test' isotope, which is used to find the values of the scaling factor $\epsilon(E)$ for the isotopes of a given element. In this connection, it should be noted that since the experimental charge radii of light (stable) nuclei and also their experimental CCCSs are generally known, it seems that the consideration of a stable isotope as a 'test' one could provide a better way of finding the scaling factor that can be used in future studies to (i) analyse CCCSs for Z $\leq$ 9 elements at a wider energy range, and (ii) provide reliable estimates for the charge-changing cross sections (and inferring charge radii), where measurements are scarce. For completeness, it is also interesting to note that although the present analysis has involved the lowest mass isotope as the 'test' one amongst the stable isotopes of a given element, except for nitrogen at $\sim$200 and 300 MeV/nucleon (for reasons indicated above), it has been verified that the use of other 'test' (stable) isotope provides the scaling factor which does not deviate more than 2$\%$ from the extracted one (Tables \ref{tab3} and \ref{tab4}), and the predicted CCCSs still provide satisfactory explanation of the experimental data involving the similar proton radii.

\section{Summary and Conclusions}
\label{sec4}

In this work, we have presented a theoretical study of the charge-changing cross sections (CCCSs) for Be, B, C, N, O, and F isotopes on a $^{12}$\rm C target in the Glauber model, involving a phenomenological scaling factor to accommodate the effect of the presence of projectile neutrons in the calculation of CCCS. The calculations have been performed in two steps: (i) In the first step, we made detailed study of the $^{28}$\rm Si + $^{12}$\rm C charge-changing cross sections at 90-1296 MeV/nucleon and obtained the energy dependence of the scaling factor in the said energy range. (ii) Based on these results, we, in the second step, proposed a method that can be used to find the scaling factor for the charge-changing cross sections of other elements. The calculations included the study of CCCSs for the isotopes of Be, B, C, N, O, and F at 200-991 MeV/nucleon. It is found that our method to find the scaling factor for the isotopes of Be, B, C, N, O, and F works well to account for the experimental data in all the cases, and the extracted scaling factor shows systematic energy dependence for the isotopes of a given element.

In conclusion, the present work seems promising in providing the energy-dependent scaling factor for Z $\leq$ 9 isotopes, needed in the study of nucleus-nucleus CCCS, involving only the projectile protons. Still we need more data on CCCS for Z $\leq$ 9 as well as other light elements, at a wider range of incident energies, in order to assess the usefulness of the scaling factor approach as adopted in this work.

\section{ Acknowledgments}

Z.H. acknowledges the UGC-BSR Research Start-Up-Grant (No.F.30-310/2016(BSR)). M.I. and Z.A.K. acknowledge the Department of Physics, Aligarh Muslim University, Aligarh for using the computational facility. A.A.U. acknowledges the Inter-University Centre for Astronomy and Astrophysics, Pune, India for support via an associateship and for hospitality.

\end{document}